\documentclass[useAMS,usenatbib]{mn2e}
\usepackage{float}
\usepackage{color}
\usepackage{graphics}
\usepackage{graphicx}
\usepackage{verbatim}

\begin{document}

\title{Capabilities of future intensity interferometers for observing fast-rotating stars: imaging with two- and three-telescope correlations}

\author[P. D. Nu\~nez, A. Domiciano de Souza]{Paul D. Nu\~nez$^{1,2,3}$\thanks{E-mail: nunez.paul@gmail.com}, A. Domiciano de Souza$^2$\\
$^{1}$Coll\`ege de France, 11 place Marcelin Berthelot 75005, Paris, France\\
$^{2}$Laboratoire Lagrange, Universit{\'e} C{\^o}te d'Azur, Observatoire de la C{\^o}te d'Azur, CNRS,\\ Bd de l'Observatoire, CS 34229, 06304 Nice cedex 4, France\\
$^{3}$Present address: JPL, Jet Propulsion Laboratory, California Institute of Technology, 4800 Oak Grove Drive, Pasadena,\\ CA 91109-8099, USA.}

\maketitle

\begin{abstract}
Future large arrays of telescopes, used as intensity interferometers, can be used to image the surfaces of stars with unprecedented angular resolution. Fast-rotating, hot stars are particularly attractive targets for intensity interferometry since shorter (blue) wavelength observations do not pose additional challenges. Starting from realistic surface brightness simulations of fast-rotating stars, we discuss the capabilities of future intensity interferometers for imaging effects such as gravity darkening and rotational deformation. We find that two-telescope intensity correlation data allow reasonably good imaging of these phenomena, but can be improved with additional higher order (e.g. three-telescope) correlation data, which contain some Fourier phase information.

\end{abstract}

\section{Introduction}

Over half a century after the success of the Narrabri Stellar Intensity Interferometer \citep{hanbury_brown0}, a future facility: the Cherenkov Telescope Array (CTA), has been proposed to be used in part as an Intensity Interferometer \citep{holder2, dainis.cta, mnras}. When used as an optical Intensity Interferometer, CTA will provide thousands of simultaneous baselines as large as a kilometer, corresponding to sub-milliarcsecond angular resolution at visible wavelengths, and thus enable to resolve the surfaces of many nearby stars. Since intensity interferometry is insensitive to atmospheric turbulence and optical imperfections, observing at short blue wavelengths, where Cherenkov telescopes are particularly sensitive, does not impose additional challenges. Also, Intensity Interferometry has been shown to be particularly sensitive to hot stars, which emit much of their radiation in the blue. Such observations would complement amplitude interferometry observations, which are typically done at longer wavelengths. Some numerical simulation studies of future intensity interferometers show that stellar surfaces can be imaged \citep{mnras, mnras2}, and recent laboratory experiments performed by \citet{dainis_nature} have resulted in the first optical images from intensity interferometry data. 

In this simulation paper we investigate the capabilities of CTA used for Stellar Intensity Interferometry (SII) observations of hot, fast-rotating stars (spectral types O, B, and A) at blue wavelengths. These stars are those presenting the highest rotation rates in the H-R diagram for non-degenerate objects. The most apparent effects of fast-rotation on the stellar surface are (1) the deformation of the outer layers of the star (rotation flattening), and (2) the latitudinal dependent emitted flux (gravity darkening), both consequences of the balance between significant centrifugal and gravitational forces within the star. Depending on the rotation rate, these two effects can lead to significant modifications in the specific intensity distribution of the star as seen by a distant observer. SII will allow performing morphological studies studies, e.g. of rotational flattening, as well as mapping the brightness (temperature) distribution on the stellar surface \citep{mnras2}. The effect of stellar rotation on the surface brightness distribution affects our understanding of radiation transport, which may depart from simple diffusion, and affect the way elements can be re-distributed within the star. Therefore, imaging at high angular resolution may in turn further our understanding of the time evolution of these stars \citep{delaa}.

The outline of this paper is the following: In Section \ref{fast-rotator_model} we introduce the model used to simulate the surface radiance distribution of fast rotating stars. In Section \ref{II} we discuss two- and three-point intensity correlations as well as their Signal-to-Noise-Ratio (SNR). In Section \ref{data_sim} we provide a detailed description of the data simulation. Simulated and laboratory image reconstructions have so far relied exclusively on two-point intensity correlation data, and one of the main goals here is to investigate the benefit of additionally using three-point correlations. In Section \ref{data_analysis} we describe our data analysis approach, focusing on model-independent image reconstruction from the simulated data, and present our results using two- and three-point intensity correlation (simulated) data. We conclude in Section \ref{conclusions}.

\section{Fast-rotating Stars} \label{fast-rotator_model}

The study of fast-rotating stars has largely benefited from the high angular resolution capabilities of modern amplitude interferometry \citep[e.g.][]{Domiciano-de-Souza2014_v569pA10, Monnier2012_v761pL3, van-Belle2012_v20p51-99}. These works showed that the photospheric structure of rapidly rotating, non-degenerate, single stars of intermediate to high masses are well described by a model based on the following assumptions: solid rotation (angular velocity $\Omega$), mass $M$ concentrated in the center of the star, and gravity darkening. From these assumptions, the stellar surface of the star is deformed (flattened) by rotation and follows the Roche equipotential (gravitational plus centrifugal),
\begin{equation} \label{eq:equipotroche}
 \Psi (\theta) =-\frac{GM}{ R(\theta)}-\frac{\Omega^{2} R^{2}(\theta) \sin ^{2}\theta }{2}=
                          -\frac{GM}{R_\mathrm{eq}}-\frac{v_\mathrm{eq}^2}{2} \ ,
\end{equation}
where $\theta$ is the co-latitude, $R_\mathrm{eq}$ and $v_\mathrm{eq}(=\Omega R_\mathrm{eq})$ are the equatorial radius and rotation velocity. Solving the Roche equipotential provides the co-latitude-dependent stellar radius $R(\theta)$. The local effective gravity is obtained directly from the gradient of Roche equipotential $\Psi$, i.e., 
\begin{equation}
g_\mathrm{eff}(\theta) =\left|-\nabla\Psi(\theta)\right| \ .
\end{equation}

Gravity darkening is included by relating $g_\mathrm{eff}(\theta)$ to the local effective temperature $T_\mathrm{eff}(\theta)$ in a generalized form of the gravity-darkening law proposed by \citet{von-Zeipel1924_v84p665-683}, i.e.,
\begin{equation}\label{gravity_darkening}
T_\mathrm{eff}(\theta)= K g_\mathrm{eff}^{\beta}(\theta) \ ,
\end{equation}
where $K$ is a proportionality constant, and $\beta$ is the gravity-darkening coefficient ($\beta=0.25$ in the original von Zeipel gravity darkening). Deviations from $\beta=0.25$ are interesting since they are indicative of convective atmospheric physics (e.g. $\beta=0.08$ \citep{lucy_1967}, or something in between \citep{zhao_2009, che_2011, Espinosa-Lara2011_v533pA43, Claret2015_v577pA87}. This model, hereafter called RVZ  (for Roche-von Zeipel) model is used in the next sections to simulate SII observations of fast-rotating stars.

\section{Intensity Interferometry}\label{II}

In a Stellar Intensity Interferometer, the star-light received at each aperture is converted into a time-dependent intensity signal with fast electronics. Time averaged correlations are found between telescope pairs (and possibly triplets, etc.) with analog correlators or off-line digital signal processing. Since the tolerance on optical imperfections/delays is set by the electronic time resolution, Intensity Interferometry can be insensitive to optical delays of the order of a meter when using detectors with a response time of a few nano-seconds.

These intensity correlations in turn provide information about the degree of coherence of light, i.e. the interferometric visibility. The time-averaged two-point intensity correlations $\langle I_i I_j\rangle$ are related to the interferometric (complex) visibility $\gamma_{ij}$ as

\begin{equation}
  \frac{\left\langle I_i I_j \right\rangle}{\left\langle I_i\right\rangle \left\langle I_j\right\rangle}=1+\left| \gamma_{ij} \right|^2 .
  \label{two-point}
\end{equation}

Since the phase of the complex visibility is not measured with two-point intensity correlations, image reconstruction from two-point correlations is non-trivial and has benefited from the use of several phase retrieval methods and image reconstruction techniques \citep{holmes, mnras, strekalov}.

However, it can be shown that higher order correlations can retain some Fourier phase information. In the case of three-point-correlations, it can be shown that (see e.g. \citep{malvimat}

\begin{equation}
    \frac{\left\langle I_i I_j I_k\right\rangle}{\left\langle I_i\right\rangle \left\langle I_j\right\rangle \left\langle I_k\right\rangle}   = 1+\left| \gamma_{ij} \right|^2+\left| \gamma_{jk} \right|^2+\left| \gamma_{ki} \right|^2+2Re\left[\gamma_{ij} \gamma_{jk} \gamma_{ki}\right].
    \label{three-point}
\end{equation}

The phase of the triple product in last term of this equation is known as the \emph{closure phase}, i.e.

\begin{equation}
  \phi_c = Arg\left[\gamma_{ij} \gamma_{jk} \gamma_{ki}\right]. \label{closure_phase}
\end{equation}

The closure phase ($\phi_c$) has the property of being insensitive to the atmospheric phase since baselines $ij$, $jk$ and $ki$ form a closed loop. Strictly speaking, three-point intensity correlations only allow measuring the cosine of the closure phase. In Section \ref{strategy} we discuss strategies for reconstructing images using two- and three-point intensity correlation measurements. While some simulated studies on image reconstruction have been performed using two-point intensity correlations, closure-phase intensity-correlation imaging is still in embryonic form. 

\subsection{Signal-to-Noise} \label{snr_section}

The signal-to-noise ($SNR_{(2)}$) for a two-point intensity correlation measurement depends on the degree of coherence $\gamma$, the area $A$ of each of the light receivers, the spectral density $n$ (number of photons per unit area per unit time, per frequency), the quantum efficiency $\alpha$, the electronic bandwidth $\Delta f$, and the observation time $t$. The explicit SNR expression can be shown to be \citep{hanbury_brown0}

\begin{equation}
SNR_{(2)}=n(\lambda, T, m_v)\,A\;\alpha\;|\gamma|^2\sqrt{\Delta f t/2}. \label{snr2}
\end{equation}

Note that $SNR_{(2)}$ is independent of the spectral bandwidth, imposing no additional difficulty of observing at individual spectral lines. A more detailed discussion of the sensitivity of CTA for SII is given by \citet{mnras}, and according to their simulations, tens of hours of integration time are required for observing bright ($m_v\sim6$) stars at blue wavelengths.

A study of the $SNR_{(3)}$ for three-point correlations has been recently performed by \citet{malvimat} and further investigated by \citet{wentz}, who find 

\begin{equation}
  SNR_{(3)}=|\gamma_{ij}\gamma_{jk}\gamma_{ki}|[ n(\lambda, T, m_v) A] ^{3/2}\Delta f \sqrt{t/\Delta \nu}, \label{snr3}
\end{equation}
where $\gamma_{ij}$ is the degree of coherence for the baseline $i-j$, $n$ is the spectral density, $A$ is the individual telescope area, $\Delta f$ is the electronic bandwidth, $t$ is the observation time, and $\Delta\nu$ is the optical bandwidth $c\Delta\lambda/\lambda^2$. Equations \ref{snr2} and \ref{snr3} will be used for adding noise to the simulated data presented in Section \ref{data_sim}.

If we assume individual CTA telescope areas of $\sim43\,\mathrm{m}^2$ (roughly $\sim3.7\,\mathrm{m}$ dishes consistent with  \citet{Bernlohr-2013}), an electronic bandwidth of $\sim2\,\mathrm{GHz}$, a quantum efficiency of $\sim60\%$, $\lambda=545\,\mathrm{nm}$, and an optical bandwidth of $\sim10^{11}\,\mathrm{s}^{-1}$ for $50\,\mathrm{hrs}$ of observing time on a $3^{rd}$ magnitude star (corresponding to $n\sim10^{-5} m^{-2} s^{-1} Hz^{-1}$, depending on stellar temperature),  we obtain $SNR_{(2)}\sim2000$ and a $SNR_{(3)}\sim3$. Both $SNR_{(2)}$ and $SNR_{(3)}$ can benefit from using multiple spectral channels since the $SNR$ improves as the square root of the number of spectral channels. In Sec. \ref{results} we estimate the number of spectral channnels needed to benefit from the additional use of three-point correlations. 

\section{Data Simulation} \label{data_sim}

\subsection{Simulation of the Pristine Images}

\begin{figure}
  \begin{center}
    \includegraphics[scale=0.45]{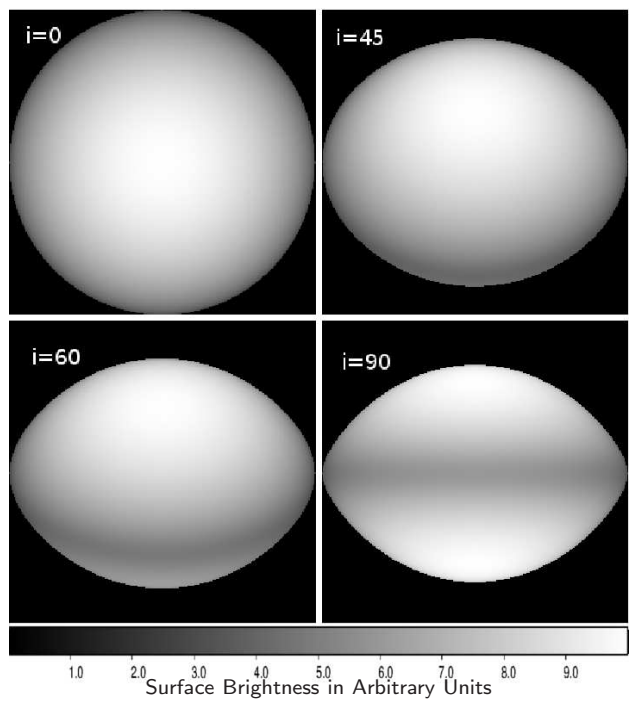}\\
    \vspace{-0.1cm}
    \small{\textsf{Surface Brightness in Arbitrary Units}}
  \end{center}
    \vspace{-0.3cm}
  \caption{\label{pristine_545} Simulated Pristine images at $545\,\mathrm{nm}$. Each image corresponds to a different inclination angle of the stellar rotation axis, where the angle is taken with respect to the line of sight. These models of fast rotators were calculated with the CHARRON code summarized in Sect.~\ref{data_sim}.}
\end{figure}

The simulation process starts by generating an angular surface brightness distribution (or intensity maps) in a certain wavelength range (see Figure \ref{pristine_545}).

These images are generated with the numerical model CHARRON (Code for High Angular Resolution of Rotating Objects in Nature) that we shortly describe below. A more detailed description is given by \citet{Domiciano-de-Souza2002_v393p345-357, Domiciano-de-Souza2012_v545pA130}. The adopted physical model for the fast-rotating star is the RVZ model summarized in Sect.~\ref{fast-rotator_model}. 

CHARRON is an IDL-based numerical model, oriented to deal with spectro-interferometric data, which means that, from the input physical parameters, it computes  monochromatic intensity maps of the stellar photosphere. From the Fourier transform (FT) of these maps one can obtain spectro-inferferometric observables (e.g. visibility amplitudes, differential phases, closure phases). Numerically, the stellar photosphere (Roche potential surface) is divided in a predefined grid with nearly identical surface area elements (typically $\sim50\,000$). From the RVZ physical equations it is possible to assign different physical quantities to each surface grid element $j$: radius $R_j$, rotation velocity $v_j (= \Omega R_j)$, effective gravity $g_{\mathrm{eff},j}$, and effective temperature $T_{\mathrm{eff},j}$.

A local specific intensity from a plane-parallel atmosphere $I_j(=I_j(g_{\mathrm{eff},j}, T_{\mathrm{eff},j}, \lambda(v_{\mathrm{proj},j}), \mu_j)$ is then associated to each surface element. The wavelength $\lambda(v_{\mathrm{proj},j})$, corresponding to the local specific intensity, is Doppler-shifted to the local rotation velocity projected onto the observer's direction $v_{\mathrm{proj},j}$. The parameter $\mu_j$ is the cosine between the normal to the surface grid element and the line-of-sight (limb darkening is thus automatically included in the model). Both $v_{\mathrm{proj},j}$ and $\mu_j$ depend on the chosen inclination $i$ of the rotation axis.

The local $I_j$ can be given by an analytic equation (including for example continuum flux, line profiles, limb darkening) or as an output of stellar atmosphere models calculated with radiative transfer codes. The results presented in this work are mainly based on $I_j$ interpolated on a pre-calculated grid of specific intensities. These pre-calculated specific intensities were obtained from the spectral synthesis code SYNSPEC \citep{Hubeny2011_SYNSPEC} and the ATLAS9 stellar atmosphere models \citep{Kurucz1979_v40p1-340}. The total intensity map of the star $I_{\lambda}$ is obtained from the juxtaposition of each $I_j$ for the apparent surface grid elements. Examples of these intensity maps (our pristine images) are given in Figure~\ref{pristine_545}. 

\begin{table}
 \caption[]{Stellar parameters adopted to create the model images shown in Fig.~\ref{pristine_545}. The adopted parameter values roughly correspond to a fast-rotating B type star.}
 \label{ta:achernar-like_parameters}
\begin{center}
\begin{tabular}{c|c}
\hline
\textbf{Model parameter}  & \textbf{Value} \\
\hline
Stellar mass $M$  		& 6 M$_\odot$ \\
Inclination angle of rotation axis $i$ 		  & 0$^\circ$,\,45$^\circ$,\,60$^\circ$,\,90$^\circ$ \\
Equatorial radius $R_\mathrm{eq}$ 		   & 11 R$_\odot$ \\
Polar radius $R_\mathrm{p}$ 	        	                &   7.83 R$_\odot$ \\
Equatorial rotation velocity $v_\mathrm{eq}$ 	&  290 km\,s$^{-1}$ \\
Gravity-darkening coefficient $\beta$ 				& 0.20 \\
Surface averaged mean temperature $\overline{T}_\mathrm{eff}$       & 15\,000\,K \\
Polar eff. temperature $T_\mathrm{p}$       & 17\,819\,K \\
Equatorial eff. temperature $T_\mathrm{eq}$& 11\,180\,K \\
\hline
\end{tabular}
\end{center}
\end{table}

\subsection{Simulation of Intensity Interferometry data}

\begin{figure}
  \hspace{-0cm}\rotatebox{-90}{\includegraphics[scale=0.45]{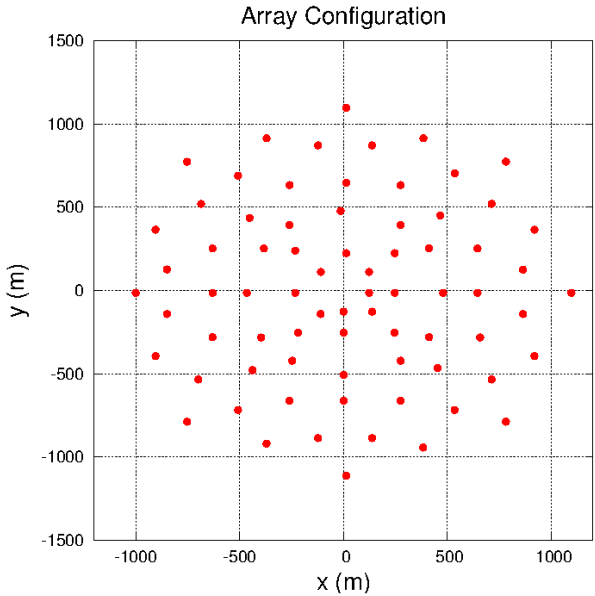}} 
  \caption{\label{cta_array} Telescope array configuration \citep{Bernlohr-2013} used for the simulations.}
\end{figure}

In order to simulate SII data, we follow the procedure described in detail by \citet{mnras}. First we set the spatial scale of the image: we concentrate on the case of a bright star of $m_v= 3$, which can have an angular diameter of $\sim0.5\,\mathrm{mas}$ as long as the temperature of the star is more than\footnote{To estimate the temperature corresponding to a particular $m_v$ and angular diameter, we use the fact that $m_v-m_0=-2.5 log(F/F_0)$, where $F_0$ is the flux corresponding to a star of $m_0$ and $F$ is the flux of our simulated star. Last, we recall that the flux is proportional to $\theta^2 T^4$, where $\theta$ is the angular diameter and $T$ is the effective temperature.} $\sim10,000\,\mathrm{K}$. Larger angular diameters than our choice of $0.5\,\mathrm{mas}$ are acceptable as long as they are not fully resolved by baselines smaller than $\sim100\,\mathrm{m}$. If needed, the pupil of the largest telescopes ($\sim24\,\mathrm{m}$, presumably located towards the center of the array) could be segmented into sub-pupils to allow observations with angular diameters much larger than $0.5\,\mathrm{mas}$. Then we take the Fast Fourier Transform (FFT) of the pristine image, and sample the data in the $(u,v)$ plane as determined by the telescope array configuration and the observing wavelength. A possible CTA array configuration \citep{Bernlohr-2013} is shown in Figure \ref{cta_array}, consisting of 74 telescopes of $4\,\mathrm{m}$ and $7\,\mathrm{m}$ telescopes, spread out over 2 kilometers. Such an array provides 2701 baselines pairs, ranging from $14\,\mathrm{m}$ to $2\,\mathrm{km}$, and 70300 telescope triplets. Two-point and three-point correlation measurements are simulated by sampling an FFT over the baselines (pairs) and triplets provided by the array, using equations \ref{two-point} and \ref{three-point}.  Gaussian noise is introduced according to equations \ref{snr2} and \ref{snr3}.

\begin{figure*}
\begin{eqnarray}
\includegraphics[scale=0.35, angle=-90]{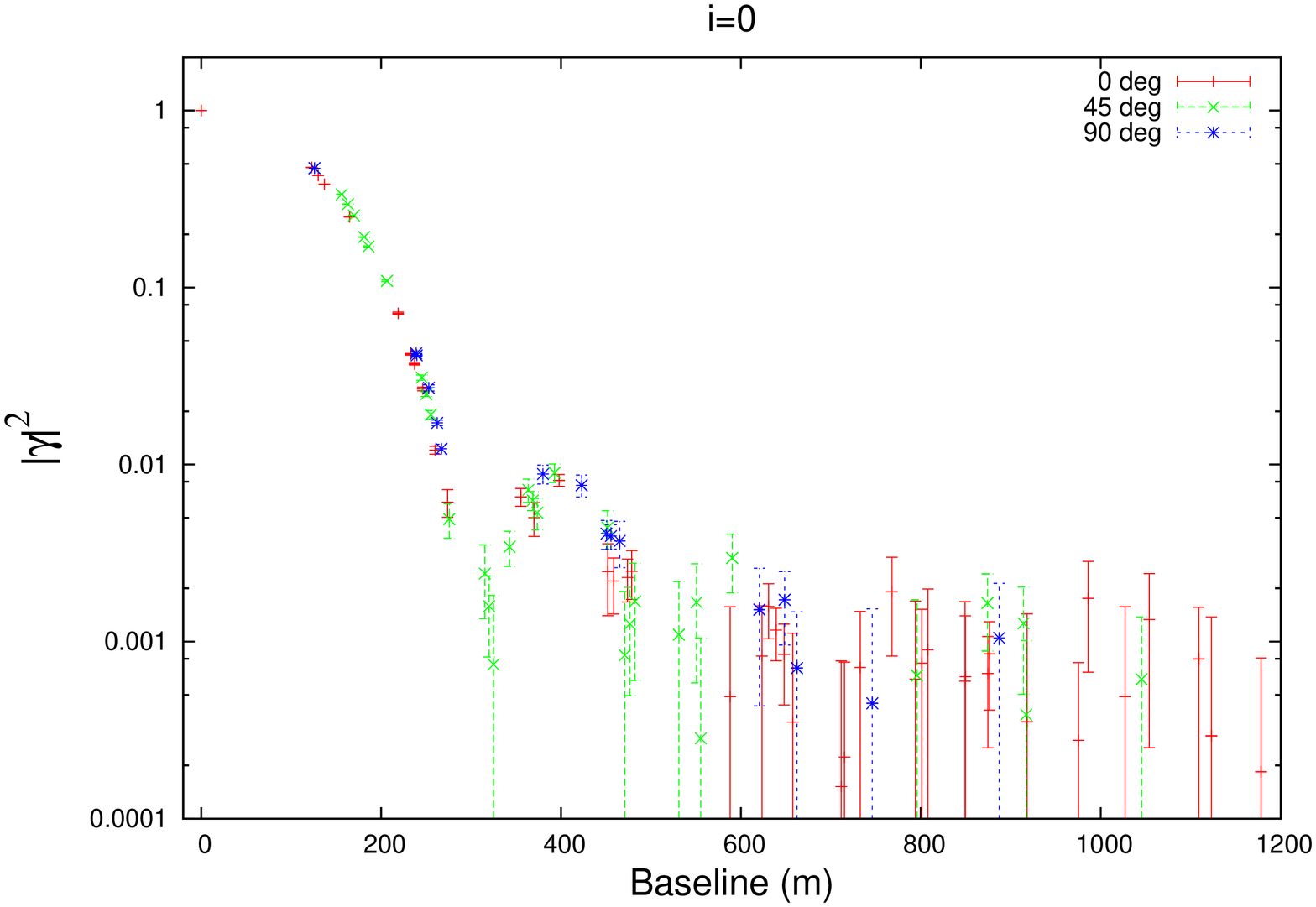} \,\,\,\,\,\,\, \includegraphics[scale=0.35, angle=-90]{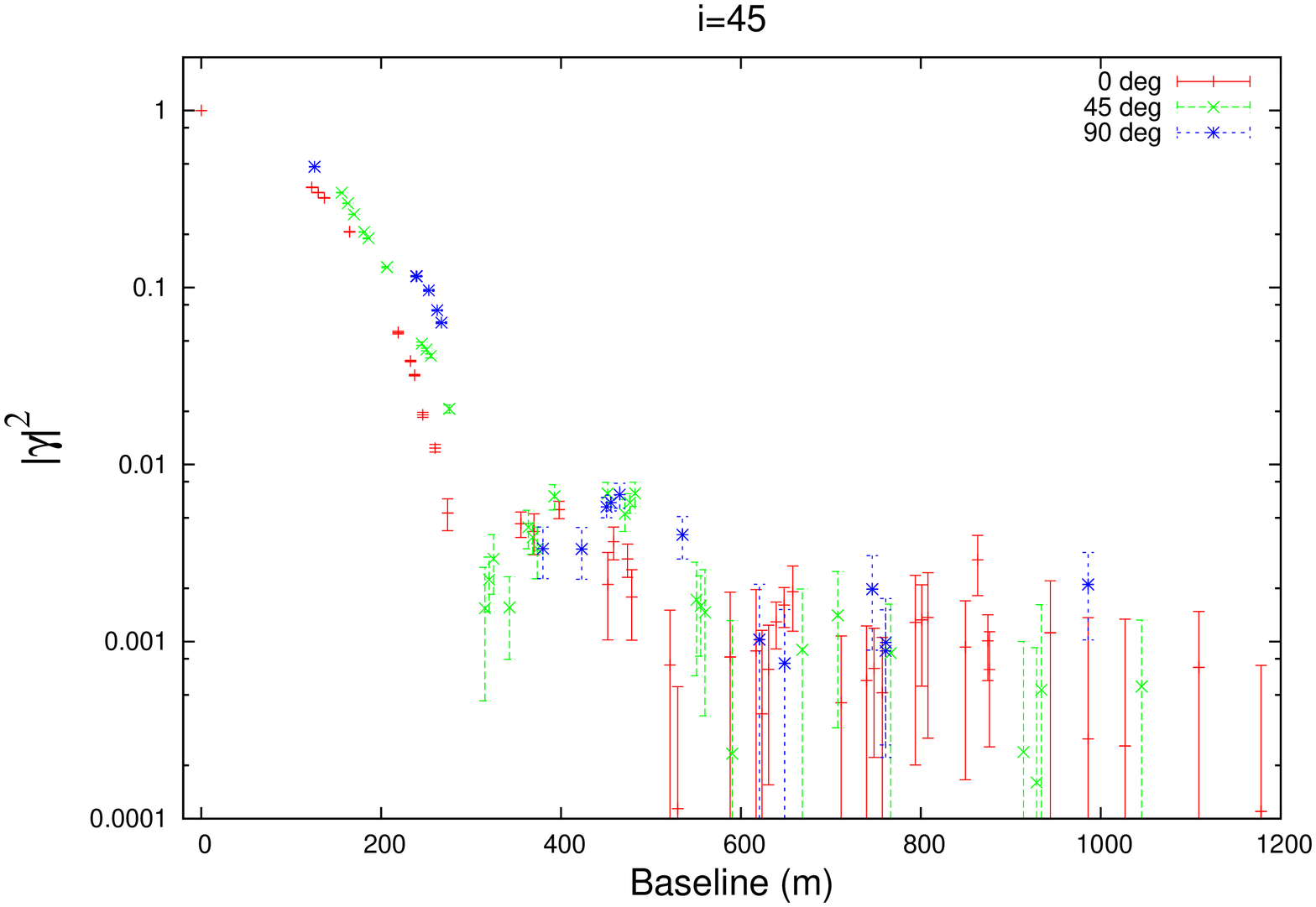}\nonumber \\\nonumber \\
\includegraphics[scale=0.35, angle=-90]{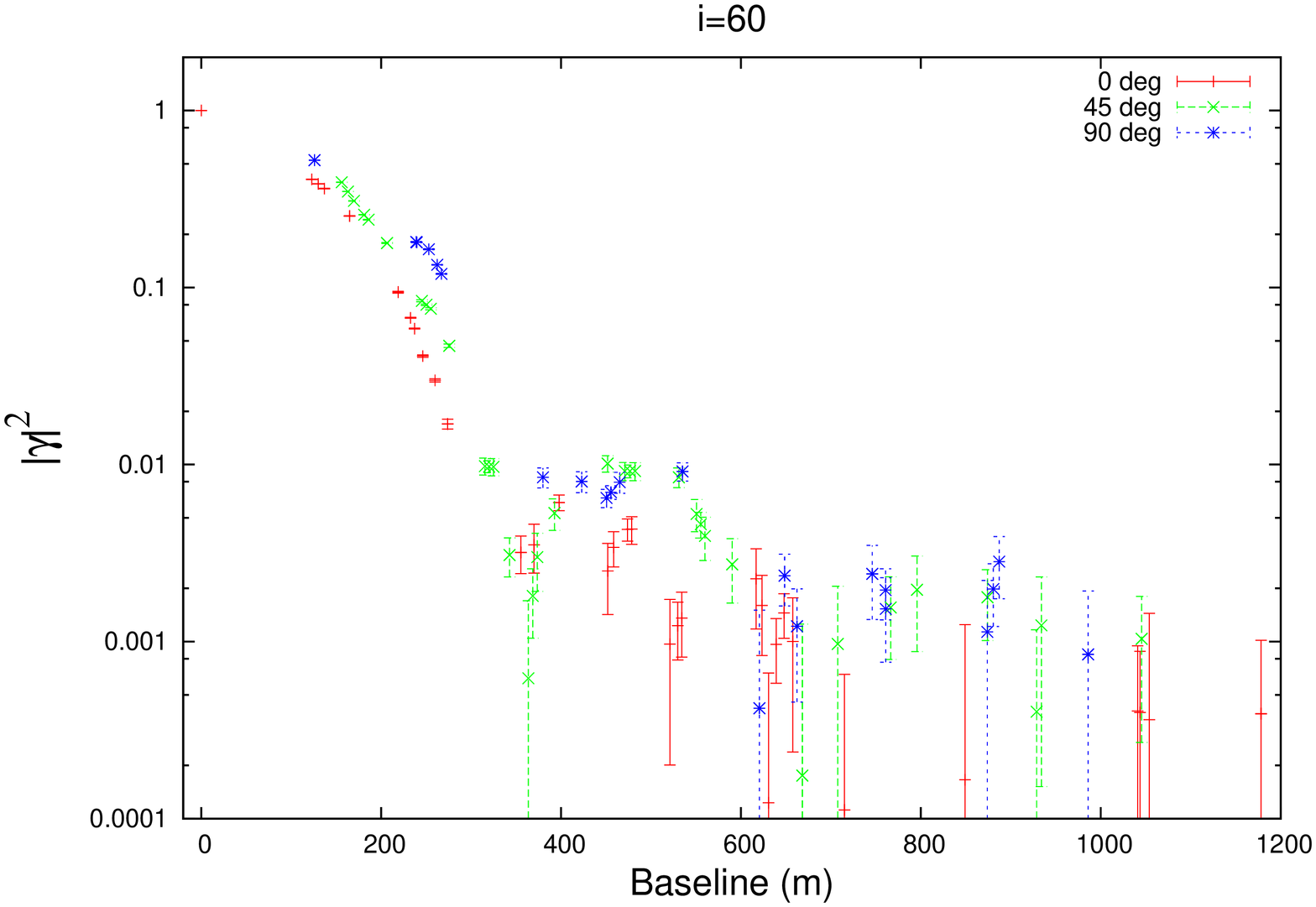} \,\,\,\,\,\,\, \includegraphics[scale=0.35, angle=-90]{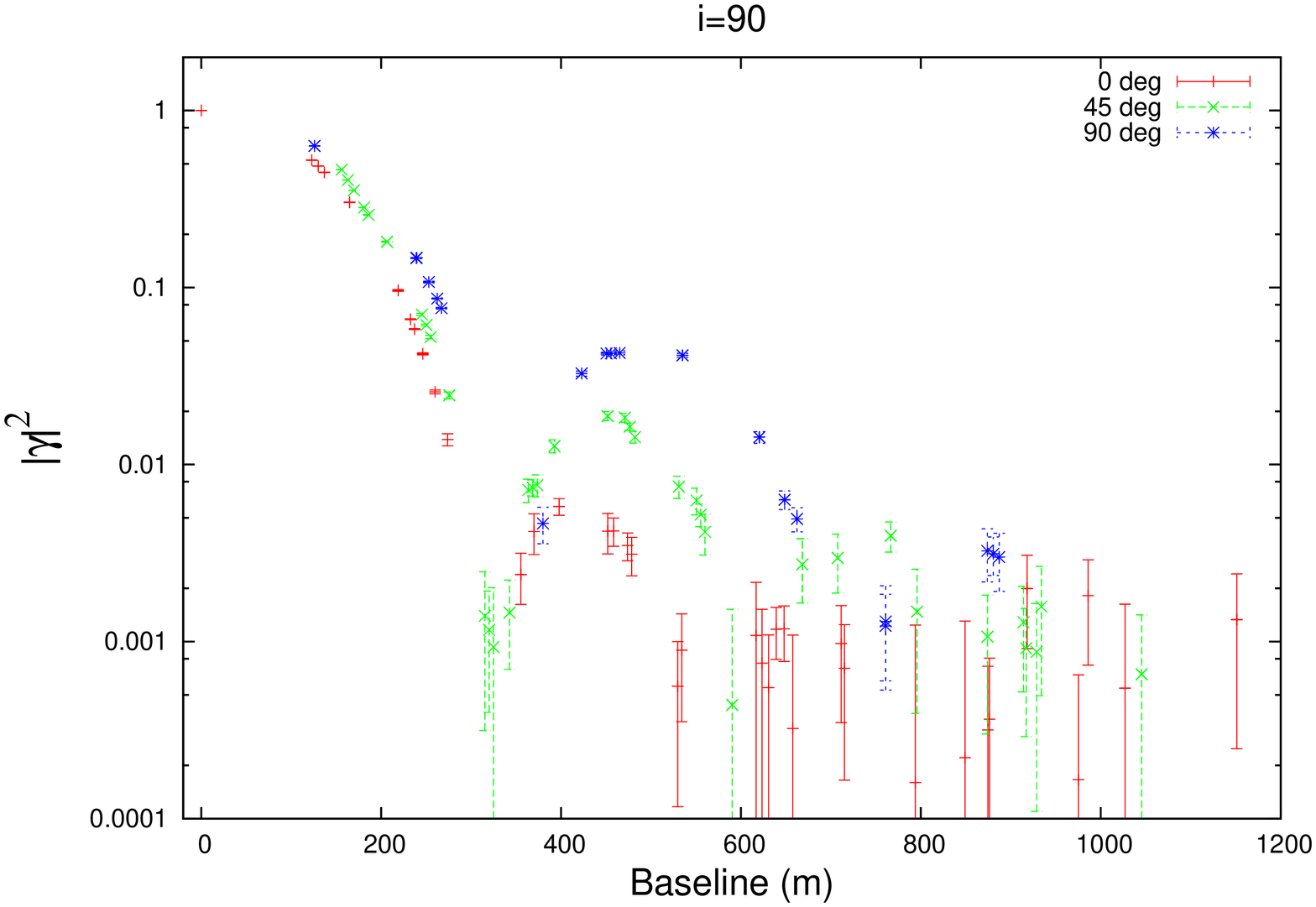} \nonumber
\end{eqnarray}
\vspace{0.5cm}
\caption{\label{data_plot} Two-point intensity correlation data at $545\,\mathrm{nm}$ for $m_v=3$ and $50\,\mathrm{hrs}$ of observation time for the different inclination angles of the fast-rotating star. The data labeled as $0\,\mathrm{deg}$ corresponds to baselines parallel to the $x$-axis in Fig. \ref{cta_array}, the other data points correspond to baselines at $45^{\circ}$ and $90^{\circ}$ with respect to the $x$-axis in Fig. \ref{cta_array}. The data are binned every $10\,\mathrm{m}$ for visual clarity.}
\end{figure*}

We simulate intensity correlation data at $545\,\mathrm{nm}$ for each of the pristine images shown in Figure \ref{pristine_545}. For simplicity we assume that the star is observed at zenith for a total of $50\,\mathrm{hrs}$, and that baselines do not drift due to earth's rotation, achieved by performing short nightly observations of, say, a few minutes. In Fig. \ref{data_plot} we show projections of the two-point intensity correlation data corresponding to $m_v=3$, using a single spectral channel. The data simulation parameters are summarized in Table \ref{sim_parameters}. To simulate data with multiple spectral channels close to each other, we simply reduce the error bars by the square root of the number of spectral channels. 

\begin{table}
 \caption[]{Data simulation parameters.}
 \label{sim_parameters}
\begin{center}
\begin{tabular}{c|c}
\hline
\textbf{Data Simulation Parameter}  & \textbf{Value} \\
\hline
Equatorial Angular Diameter  & $0.5\,\mathrm{mas}$\\
Apparent Visual Magnitude & 3 \\
Observation Time & $50\,\mathrm{hrs}$\\
Wavelength & $545\,\mathrm{nm}$ \\
Optical Bandwidth per spectral channel & $10^{11}\,\mathrm{s}^{-1}$ \\

\hline
\end{tabular}
\end{center}
\end{table}

\section{Data Analysis and Interpretation} \label{data_analysis}

As in amplitude interferometry, the two-point intensity correlations provide physical information on fast-rotators, such as equatorial angular diameter and rotation flattening. As shown in Figure \ref{data_plot} the high spatial frequencies attained by SII allow to probe the second and third visibility lobes, which are more sensitive to limb and gravity darkening than the first lobe alone. The three-point correlation data can be sensitive to asymmetries in the stellar brightness distribution, mainly due to gravity darkening in the case of fast rotating stars.

A commonly adopted strategy to extract the physical information from the data is to apply a model fitting procedure that allows to estimate some model parameters and their uncertainties. For modern amplitude interferometers the uncertainties on the model parameters range typically of the order of $\simeq1\%-10\%$, depending on the parameter \citep[e.g.][]{Domiciano-de-Souza2014_v569pA10, Monnier2012_v761pL3}. Since the $(u,v)$ coverage of CTA can be so dense, we rather adopt model-independent image reconstruction strategies described in Sec. \ref{strategy}. An important point we will investigate is the imaging benefit of using three-point correlations in addition to two-point correlation data.

\subsection{Image reconstruction strategy} \label{strategy}

We test model-independent image reconstruction tools to analyze the data displayed in Figure \ref{data_plot}. We first follow the strategy described in detail by \citep{mnras2}: The first step consist in estimating the phase from Fourier magnitude data alone, using the ``Cauchy-Riemann'' phase reconstruction method, based on the theory of analytic functions \citep{holmes, mnras}. The main idea behind the Cauchy-Riemann method is that the Fourier Transform of an object of finite extent is an analytic function. Therefore, one can use the Cauchy-Riemann equations (in polar coordinates) to relate the Fourier magnitude to the phase. Since the Cauchy-Riemann algorithm relies on the calculation of derivatives, an analytic function is fitted to the data as described in \citet{mnras}. 

The first estimate can be further improved by using two different algorithms:  the ``Gerchberg-Saxton'' algorithm \citep{gs}, which consists in going ``back and forth'' between image and Fourier space, each time imposing general restrictions in each domain. In our case, the ``Gerchberg-Saxton'' restrictions in image space consist in setting setting pixels outside a circular region to zero, where the size of the circular region is determined by the smallest closest zero of the Fourier Magnitude. The restriction in Fourier space is that the Modulus of the Fourier transform of the reconstructed image should be replaced by the analytic function fit. We note that a poor fit can already introduce artifacts in the reconstructed image.

The reconstructed image can be further improved with yet another algorithm: the ``Multi-aperture Image Reconstruction Algorithm'' (MiRA) \citep{thiebaut}: an optimization algorithm which maximizes agreement between the Fourier transform of the reconstructed image and the data. The MiRA is particularly good for removing artifacts that may have arisen from the initial fit of the data. The reason for using a seemingly complex chain of algorithms, and not a single algorithm, has been justified empirically via simulations performed by \citet{mnras2}, and the different algorithms seem to complement each-other. In the context of this paper, the MiRA is also particularly useful for using the three-point correlation (closure-phase) data, and to test if such higher-order correlation measurements are indeed useful for image reconstruction.

Recall from eq. \ref{three-point} that a three-point intensity correlation allows measuring the cosine of the triple product of the degrees of coherence of in a closed loop of baselines. The phase of the triple product (eq. \ref{closure_phase}) is the closure phase known in amplitude interferometry, but only the cosine of the closure phase is accessible with intensity interferometry measurements. This results in having a sign (and modulo $2\pi$) ambiguity in the closure phase measurement. Provided sufficient SNR, the sign ambiguity should be removable by performing observations at two (or more) close wavelengths separated by $\Delta\lambda$, and estimating the change in the three-point correlation for a given $\Delta\lambda$. Measuring three point correlations at two close wavelengths yields two closure phase equations of the form of Equation \ref{three-point}. If the object's surface brightness distribution does not differ largely between the two different wavelengths, the two closure phase equations should mainly differ by an additive term proportional to the ratio of the two wavelengths (in a first order approximation). This should allow removing the sign ambiguity by choosing the sign of the closure phase from one of the two closure phase equations, and then checking if it is consistent with the other equation. 

The problem of removing the sign ambiguity in the closure phase and how it affects the SNR deserves further investigation and will be treated in a future paper. In the simulations presented here, we assume that the sign ambiguity has already been removed with the approach described above, and we have not included this phase degeneracy removal in our simulations\footnote{All the codes used for our simulations are available upon request.}. The main interest is to see how three-point correlations with (much) lower SNR than two-point correlations can improve image reconstructions. 

\subsection{Image reconstruction results} \label{results}

\begin{figure}
  \begin{center}
    \includegraphics[scale=0.3]{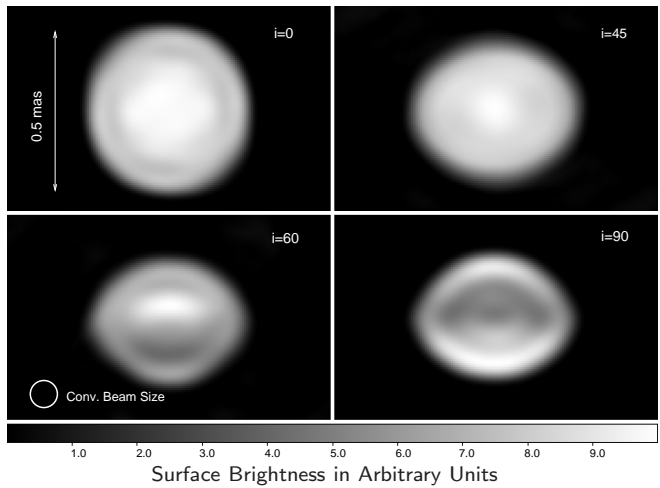}\\
    \vspace{-0.2cm}
    \textsf{Surface Brightness in Arbitrary Units}\\
  \end{center}
  \vspace{-0.3cm}
  \caption{\label{rec_no_phase} Image reconstructions without three-point correlation information.}
\end{figure}

In this section we compare image reconstructions using two-point correlation data with reconstructions using additional three-point correlation data. We first investigate image reconstruction quality when using only two-point correlation data for a single spectral channel, and find reasonably good image reconstructions as shown in Figure \ref{rec_no_phase}, which may already allow to infer the rotational flattening and the inclination angle in some cases. To quantify image reconstruction quality, we use the Mean Square Error (MSE) between the pristine image and the reconstructed image. In order to be more certain that we can correctly determine the stellar parameters from an image reconstruction, we compare each reconstruction in Figure \ref{rec_no_phase} with all the different pristine images in Figure \ref{pristine_545}, and provide the values of the MSE between each reconstructed images and pristine images in Table \ref{table_wo_phase}. For each reconstruction, the lowest MSE should be found for its corresponding pristine image. As can be seen in the Table, the reconstruction for $i=45^{\circ}$ is more similar to the pristine image with $i=0^\circ$, i.e. we have reconstructed at least one image for which we cannot determine the correct inclination angle and gravity darkening.

\begin{table}
  \caption{\label{table_wo_phase}Mean Square Error quantifying the difference between the pristine images (left column) and the reconstructed images (top row) \emph{without closure phase data}, using a single spectral channel. The images used for computing the Mean Square Error are shown in Figures \ref{pristine_545} and \ref{rec_no_phase}. The minimum MSE of each column is shown in bold-face. Note that the reconstruction of the case of $i=45$ can be confused with a pristine image with $i=0$. }
  \hspace{-0.5cm}
  \scalebox{0.82}{
    \begin{tabular}{||l ||c|c|c|c||}
      \hline
      Prist\textbackslash  Rec  & i=0        &    i=45    &     i=60      &      i=90 \\\hline
      i=0       &   $\mathbf{1.37\times 10^{-3}}$ & $\mathbf{1.24\times 10^{-3}}$  &  $5.50\times 10^{-3}$            &  $5.39\times 10^{-3}$ \\\hline
      i=45      &   $6.44\times 10^{-3}$          & $1.43\times 10^{-3}$           &  $1.43\times 10^{-3}$            &  $2.00\times 10^{-3}$ \\\hline
      i=60      &   $1.24 \times 10^{-2}$         & $5.25\times 10^{-3}$           &  $\mathbf{8.43\times 10^{-4}}$   &  $2.40\times 10^{-3}$ \\\hline
      i=90      &   $9.92\times 10^{-3}$          & $3.36\times 10^{-3}$           &  $1.92\times 10^{-3}$            &  $\mathbf{9.20\times 10^{-4}}$ \\\hline
    \end{tabular}
  }
\end{table}

\begin{table}
  \caption{\label{table_w_phase}Mean Square Error between the pristine images (left column) and the reconstructed images (top row) \emph{with closure phase data}, using 1000 spectral channels. The images used for computing the Mean Square Error are shown in Figures \ref{pristine_545} and \ref{rec_phase}. The minimum MSE of each column is shown in bold-face. Note that it is unlikely to confuse a reconstruction with the incorrect pristine image.}
  \hspace{-0.5cm}
  \scalebox{0.82}{
    \begin{tabular}{||l ||c|c|c|c||}
      \hline
      Prist \textbackslash Rec  & i=0        &    i=45    &     i=60      &      i=90 \\\hline
      i=0       &   $\mathbf{1.17\times 10^{-3}}$ & $1.44\times 10^{-3}$           &  $4.62\times 10^{-3}$           &  $5.67\times 10^{-3}$ \\\hline
      i=45      &   $2.32\times 10^{-3}$          & $\mathbf{1.01\times 10^{-3}}$  &  $4.47\times 10^{-4}$           &  $2.32\times 10^{-3}$ \\\hline
      i=60      &   $1.15 \times 10^{-2}$         & $4.27\times 10^{-3}$           &  $\mathbf{7.16\times 10^{-4}}$  &  $2.97\times 10^{-3}$ \\\hline
      i=90      &   $9.24\times 10^{-3}$          & $3.16\times 10^{-3}$           &  $1.60\times 10^{-3}$           &  $\mathbf{1.02\times 10^{-3}}$ \\\hline
    \end{tabular}
  } 
\end{table}

\begin{figure}
  \begin{center}
    \includegraphics[scale=0.3]{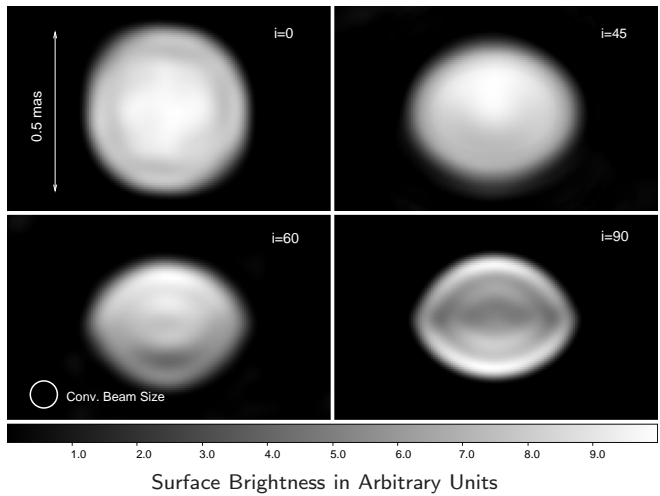}\\
    \vspace{-0.1cm}
    \textsf{Surface Brightness in Arbitrary Units}\\
  \end{center}
    \vspace{-0.2cm}
  \caption{\label{rec_phase} Image reconstructions using three-point correlation information.}
\end{figure}

Next we investigate whether or not three-point correlation data (in addition to two-point correlations) over a single spectral channel improve image reconstructions. We showed in Sec. \ref{snr_section} that $SNR_{(3)}\sim3$ for a single spectral channel, so in principle, we should be able to see improvements when using three-point correlation data. We also expected image reconstructions to improve due to the much larger number of telescope triplets than pairs, but we do not find a significant improvement using the analysis method presented in this paper, at least for the case of this object observed with this array configuration. Clearly, there is still room for improvement on the data analysis method.

\begin{figure}
  \begin{center}
  \includegraphics[scale=0.32, angle=-90]{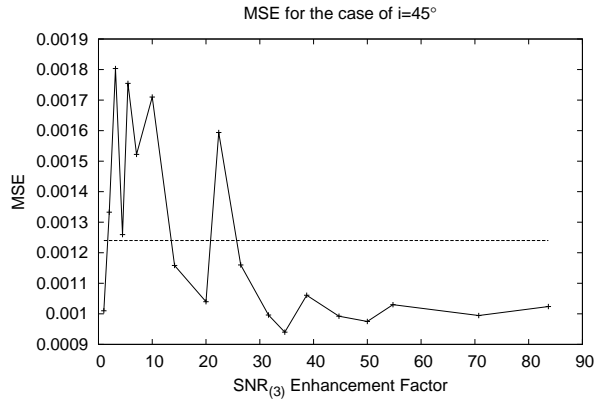}
  \end{center}
  \vspace{0.3cm}
  \caption{\label{snr_enhanced}Mean Square Error for the image reconstruction corresponding to the case $i=45^{\circ}$ with its corresponding pristine image as a function of the enhancement factor for $SNR_{(3)}$. The horizontal line is the lowest MSE found in the second column of Table \ref{table_wo_phase}. A reliable improvement in the image reconstruction can be seen for an enhancement factor approximately greater than $\sim30$.}
\end{figure}

We now consider methods for increasing the SNR, especially for three-point correlations. The amount by which the $SNR_{(3)}$ must be enhaced is empirically estimated from the (problematic) example of $i=45^{\circ}$, whose reconstruction is more similar to a pristine image with $i=0^{\circ}$ with a corresponding $MSE=1.24\times10^{-3}$. As shown in Figure \ref{snr_enhanced}, we have found empirically that $SNR_{(3)}$ needs to be increased by a factor of $\sim30$ (corresponding to $SNR_{(3)}\sim90$) in order to see a noticeable improvement in the image reconstruction quality.

One alternative to increase the SNR is to simply use much faster detectors, but even if much faster detectors existed, their use would require the optical precision comparable to amplitude interferometers. We have also considered increasing the telescope area, but the telescope area cannot increase much more before individual telescopes would start to resolve the source, resulting in a smaller correlation signal \citep{janvida}. 

Last we investigate the improvement of image reconstructions when using multiple spectral channels. For the fast-rotating star studied in this paper ($m_v=3$, $50\,\mathrm{hrs}$), we have found that at least $\sim1000$ spectral channels would need to be used in order to see a noticeable improvement in image reconstructions. When 1000 spectral channels are used with three-point correlations, we also find that reconstructed images are most similar to their corresponding pristine images, as shown in Table \ref{table_w_phase}. In other words, Table \ref{table_w_phase} shows that the minimum MSE on the diagonal, unlike Table \ref{table_wo_phase}. Our lower limit of ($\sim1000$) required spectral channels is based on the assumption that the stellar image does not change much within the optical bandwidth, which is reasonable if we avoid observing over spectral lines. The required number of spectral channels may also be different for another telescope array configuration, e.g. one containing many redundant telescope pairs and triplets.

\section{Discussion and conclusions} \label{conclusions}

Fast rotation is a common characteristic of hot, massive stars. It influences the structure and evolution during the whole lifetime of these stars, governing the loss of radiation, mass, and angular momentum. The circumstellar environment of fast-rotators is thus intimately connected to the angular momentum in the central object. In the stellar interior, rotation induces circulations of mass, providing additional mixing of material that changes the stellar structure, lifetime, production of chemical elements. The final stages of massive stars is also dependent on the previous evolution history, where rotation plays an important role.

In the present work we demonstrate the capacity of SII to resolve the apparent intensity distribution of fast-rotating stars allowing to reconstruct high angular resolution images that directly reveal, independently of any adopted model, their position angle, apparent rotation flattening, and gravity darkening (polar to equatorial intensity distribution).

The use of model fitting procedure can in addition be adopted to complement the reconstruction imaging approach in order to estimate several stellar physical parameters and their uncertainties.

The capacity of high angular resolution techniques to study fast-rotating stars was proven by recent results on image reconstruction and model fitting obtained from amplitude interferometry in the last decade for the few brightest (and thus closest) fast-rotating stars, which amount to half a dozen objects. 

The very high angular resolution attained by SII can take us one step forward allowing a more complete and systematic study of stellar rotation effects on a much larger number of targets. For example, considering SII observations at CTA with baselines up to $1000\,\mathrm{m}$ at wavelengths of $\sim500\,\mathrm{nm}$ allow to measure at least the three first visibility lobes of stars having angular diameters as small as $\sim0.5\,\mathrm{mas}$. If the image reconstruction strategy presented here is adopted, we showed that for fast rotators brighter than $m_v=3$ a very large number ($\sim1000$) of spectral channels are needed in order for three-point correlations to be beneficial. However, improvements in the data analysis method may relax this requirement on the number of spectral channels. Even with these requirements, the number of possible targets for SII imaging studies of rapid rotators given in the catalogue of \citet{van-Belle2012_v20p51-99} is 35 (5 times more than the present number of observed fast-rotators). If one relaxes the $m_v=3$ constraint then many more targets (nearly all the 354 stars in the van Belle's catalogue, with roughly half this number per hemisphere) can be investigated by SII with simply two-point correlations (with the associated possible complications shown in this work). Thus, compared to the performance of present amplitude interferometers, SII can potentially increase by a factor $\sim20-30$ the number of observable targets for studies of fast-rotation across the H-R diagram.

Finally, we note that although in this paper we concentrated on hot, massive, fast-rotating stars, which is one of the most obvious science cases for SII with CTA, there are several other targets (e.g. binary stars, stars with photospheric spots from different origins) that are bright ($m_v<3$) and may have even stronger three-point correlation signals.

\section{Acknowledgements}

We would like to thank the referee (Gerard van Belle) for his helpful comments and revision of our manuscript.

\bibliography{sii_bibliography}

\begin{thebibliography}{28}
\expandafter\ifx\csname natexlab\endcsname\relax\def\natexlab#1{#1}\fi

\bibitem[{{Bernl{\"o}hr} {et~al}\mbox{.}(2013){Bernl{\"o}hr}, {Barnacka},
  {Becherini}, {Blanch Bigas}, {Carmona}, {Colin}, {Decerprit}, {Di Pierro},
  {Dubois}, {Farnier}, {Funk}, {Hermann}, {Hinton}, {Humensky}, {Kh{\'e}lifi},
  {Kihm}, {Komin}, {Lenain}, {Maier}, {Mazin}, {Medina}, {Moralejo}, {Nolan},
  {Ohm}, {de O{\~n}a Wilhelmi}, {Parsons}, {Paz Arribas}, {Pedaletti}, {Pita},
  {Prokoph}, {Rulten}, {Schwanke}, {Shayduk}, {Stamatescu}, {Vallania},
  {Vorobiov}, {Wischnewski}, {Yoshikoshi}, {Zech}, \& {CTA
  Consortium}}]{Bernlohr-2013}
{Bernl{\"o}hr} K. {et~al.}, 2013, Astroparticle Physics, 43, 171

\bibitem[{{Che} {et~al}\mbox{.}(2011){Che}, {Monnier}, {Zhao}, {Pedretti},
  {Thureau}, {M{\'e}rand}, {ten Brummelaar}, {McAlister}, {Ridgway}, {Turner},
  {Sturmann}, \& {Sturmann}}]{che_2011}
{Che} X. {et~al.}, 2011, ApJ, 732, 68

\bibitem[{{Claret}(2015)}]{Claret2015_v577pA87}
{Claret} A., 2015, A\& A, 577, A87

\bibitem[{{Delaa} {et~al}\mbox{.}(2013){Delaa}, {Zorec}, {Domiciano de Souza},
  {Mourard}, {Perraut}, {Stee}, {Fr{\'e}mat}, {Monnier}, {Kraus}, {Che},
  {B{\'e}rio}, {Bonneau}, {Clausse}, {Challouf}, {Ligi}, {Meilland},
  {Nardetto}, {Spang}, {McAlister}, {ten Brummelaar}, {Sturmann}, {Sturmann},
  {Turner}, {Farrington}, \& {Goldfinger}}]{delaa}
{Delaa} O. {et~al.}, 2013, A\& A, 555, A100

\bibitem[{{Domiciano de Souza} {et~al}\mbox{.}(2012){Domiciano de Souza},
  {Hadjara}, {Vakili}, {Bendjoya}, {Millour}, {Abe}, {Carciofi}, {Faes},
  {Kervella}, {Lagarde}, {Marconi}, {Monin}, {Niccolini}, {Petrov}, \&
  {Weigelt}}]{Domiciano-de-Souza2012_v545pA130}
{Domiciano de Souza} A. {et~al.}, 2012, A\&A, 545, A130

\bibitem[{{Domiciano de Souza} {et~al}\mbox{.}(2014){Domiciano de Souza},
  {Kervella, P.}, {Moser Faes, D.}, {Dalla Vedova, G.}, {M{\'e}rand, A.}, {Le
  Bouquin, J.-B.}, {Espinosa Lara, F.}, {Rieutord, M.}, {Bendjoya, P.},
  {Carciofi, A. C.}, {Hadjara, M.}, {Millour, F.}, \& {Vakili,
  F.}}]{Domiciano-de-Souza2014_v569pA10}
{Domiciano de Souza} A. {et~al.}, 2014, A\&A, 569, A10

\bibitem[{{Domiciano de Souza} {et~al}\mbox{.}(2002){Domiciano de Souza},
  {Vakili}, {Jankov}, {Janot-Pacheco}, \&
  {Abe}}]{Domiciano-de-Souza2002_v393p345-357}
{Domiciano de Souza} A., {Vakili} F., {Jankov} S., {Janot-Pacheco} E., {Abe}
  L., 2002, A\&A, 393, 345

\bibitem[{{Dravins}, {Lagadec} \& {Nu{\~n}ez}(2015){Dravins}, {Lagadec}, \&
  {Nu{\~n}ez}}]{dainis_nature}
{Dravins} D., {Lagadec} T., {Nu{\~n}ez} P.~D., 2015, Nat Commun, 6

\bibitem[{{Dravins} {et~al}\mbox{.}(2013){Dravins}, {LeBohec}, {Jensen},
  {Nu{\~n}ez}, \& {CTA Consortium}}]{dainis.cta}
{Dravins} D., {LeBohec} S., {Jensen} H., {Nu{\~n}ez} P.~D., {CTA Consortium},
  2013, Astroparticle Physics, 43, 331

\bibitem[{{Espinosa Lara} \& {Rieutord}(2011)}]{Espinosa-Lara2011_v533pA43}
{Espinosa Lara} F., {Rieutord} M., 2011, A\& A, 533, A43

\bibitem[{Gerchberg(1972)}]{gs}
Gerchberg, R. W. \&~Saxton W.~O., 1972, Optik, 35, 237

\bibitem[{{Hanbury Brown}(1974)}]{hanbury_brown0}
{Hanbury Brown} R., 1974, {The intensity interferometer. Its applications to
  astronomy}, {Hanbury Brown, R.}, ed.

\bibitem[{{Holmes} \& {Belen'kii}(2004)}]{holmes}
{Holmes} R.~B., {Belen'kii} M.~S., 2004, Journal of the Optical Society of
  America A, 21, 697

\bibitem[{{Hubeny} \& {Lanz}(2011)}]{Hubeny2011_SYNSPEC}
{Hubeny} I., {Lanz} T., 2011, {Synspec: General Spectrum Synthesis Program}

\bibitem[{{Kurucz}(1979)}]{Kurucz1979_v40p1-340}
{Kurucz} R.~L., 1979, ApJS, 40, 1

\bibitem[{{Le Bohec} \& {Holder}(2006)}]{holder2}
{Le Bohec} S., {Holder} J., 2006, ApJ, 649, 399

\bibitem[{{Lucy}(1967)}]{lucy_1967}
{Lucy} L.~B., 1967, Zeitschrift für Astrophysik, 65, 89

\bibitem[{{Malvimat}, {Wucknitz} \& {Saha}(2014){Malvimat}, {Wucknitz}, \&
  {Saha}}]{malvimat}
{Malvimat} V., {Wucknitz} O., {Saha} P., 2014, MNRAS, 437, 798

\bibitem[{{Monnier} {et~al}\mbox{.}(2012){Monnier}, {Che}, {Zhao},
  {Ekstr{\"o}m}, {Maestro}, {Aufdenberg}, {Baron}, {Georgy}, {Kraus},
  {McAlister}, {Pedretti}, {Ridgway}, {Sturmann}, {Sturmann}, {ten Brummelaar},
  {Thureau}, {Turner}, \& {Tuthill}}]{Monnier2012_v761pL3}
{Monnier} J.~D. {et~al.}, 2012, ApJL, 761, L3

\bibitem[{{Nu{\~n}ez} {et~al}\mbox{.}(2012{\natexlab{a}}){Nu{\~n}ez}, {Holmes},
  {Kieda}, \& {Lebohec}}]{mnras}
{Nu{\~n}ez} P.~D., {Holmes} R., {Kieda} D., {Lebohec} S., 2012{\natexlab{a}},
  MNRAS, 419, 172

\bibitem[{{Nu{\~n}ez} {et~al}\mbox{.}(2012{\natexlab{b}}){Nu{\~n}ez}, {Holmes},
  {Kieda}, {Rou}, \& {LeBohec}}]{mnras2}
{Nu{\~n}ez} P.~D., {Holmes} R., {Kieda} D., {Rou} J., {LeBohec} S.,
  2012{\natexlab{b}}, MNRAS, 424, 1006

\bibitem[{{Rou} {et~al}\mbox{.}(2013){Rou}, {Nu{\~n}ez}, {Kieda}, \&
  {LeBohec}}]{janvida}
{Rou} J., {Nu{\~n}ez} P.~D., {Kieda} D., {LeBohec} S., 2013, MNRAS, 430, 3187

\bibitem[{{Strekalov}, {Kulikov} \& {Yu}(2014){Strekalov}, {Kulikov}, \&
  {Yu}}]{strekalov}
{Strekalov} D.~V., {Kulikov} I., {Yu} N., 2014, Optics Express, 22, 12339

\bibitem[{{Thi{\'e}baut}(2009)}]{thiebaut}
{Thi{\'e}baut} E., 2009, New Astronomy Reviews, 53, 312

\bibitem[{{van Belle}(2012)}]{van-Belle2012_v20p51-99}
{van Belle} G.~T., 2012, A\&AR, 20, 51

\bibitem[{{von Zeipel}(1924)}]{von-Zeipel1924_v84p665-683}
{von Zeipel} H., 1924, MNRAS, 84, 665

\bibitem[{{Wentz} \& {Saha}(2015)}]{wentz}
{Wentz} T., {Saha} P., 2015, MNRAS, 446, 2065

\bibitem[{{Zhao} {et~al}\mbox{.}(2009){Zhao}, {Monnier}, {Pedretti}, {Thureau},
  {M{\'e}rand}, {ten Brummelaar}, {McAlister}, {Ridgway}, {Turner}, {Sturmann},
  {Sturmann}, {Goldfinger}, \& {Farrington}}]{zhao_2009}
{Zhao} M. {et~al.}, 2009, ApJ, 701, 209

\end{thebibliography}

\end{document}